\documentclass{revtex4-1}
\usepackage[T1]{fontenc}
\usepackage[utf8]{inputenc}

\usepackage[english]{babel}
\usepackage{graphicx}

\usepackage{amsmath}
\usepackage{amsfonts}

\usepackage[unicode=true,pdfusetitle,
 bookmarks=true,bookmarksnumbered=false,bookmarksopen=false,
 breaklinks=false,pdfborder={0 0 1},backref=false,colorlinks=true,
 linkcolor=black,citecolor=black,urlcolor=blue,pdfauthor={Feliks Nüske}]
 {hyperref} 

\begin{document}
\title{Coarse-graining Molecular Systems by Spectral Matching}

\author{Feliks Nüske}
\email{feliks.nueske@rice.edu}

\author{Lorenzo Boninsegna}

\author{Cecilia Clementi}
\email{cecilia@rice.edu}

\affiliation{Center for Theoretical Biological Physics, and Department of Chemistry, Rice University}

\begin{abstract}
Coarse-graining has become an area of tremendous importance within many
different research fields. For molecular simulation, coarse-graining bears the
promise of finding simplified models such that long-time simulations of
large-scale systems become computationally tractable. While significant
progress has been made in tuning thermodynamic properties of reduced models, it
remains a key challenge to ensure that relevant kinetic properties are retained
by coarse-grained dynamical systems. In this study, we focus on data-driven
methods to preserve the rare-event kinetics of the original system, and make
use of their close connection to the low-lying spectrum of the system's
generator. Building on work by \textit{Crommelin and Vanden-Eijnden, SIAM
Multiscale Model. Simul. (2011)}, we present a general framework, called
spectral matching, which directly targets the generator's leading eigenvalue
equations when learning parameters for coarse-grained models. We discuss
different parametric models for effective dynamics and derive the resulting
data-based regression problems. We show that spectral matching can be used to
learn effective potentials which retain the slow dynamics, but also to correct
the dynamics induced by existing techniques, such as force matching.
\end{abstract}

\maketitle

\section{Introduction}
\textit{Coarse-graining} or \textit{model reduction} is the process of
describing a high-dimensional and complex dynamical system by a smaller set of
variables, and of providing a new set of governing equations for this reduced
description. Coarse-graining has become a fundamental challenge in many
different areas of science, such as finance, atmospheric science, or molecular
biology. Two of the central reasons for the importance of coarse-graining are
that, firstly, analysis or numerical simulation of high-dimensional systems is
often challenging or simply infeasible, and secondly, not all detailed features
of the full system are needed in order to answer questions of scientific
interest.
For molecular systems, important contributions to the field include
coarse-graining in structural space, where the physical representation of a
system is simplified.
Several approaches have been proposed to design coarse-grained models for large
molecular systems that either reproduce structural features of fine-grained
(atomistic) models (bottom-up)
\cite{Reith2003,izvekov2005multiscale,noid2008multiscale,Shell2008,Wang2009,Noid2013} 
or reproduce experimentally measured properties
for one or a range of systems (top-down) 
\cite{Lyubartsev1995,Nielsen2003,Marrink2004,Marrink2007,Shinoda2007,Monticelli2008,ClementiCOSB}. 

An alternative approach is coarse-graining in configurational space, where a
transformation of variables is applied to arrive at a smaller set of
descriptors. Notable examples along these lines are the Mori--Zwanzig
formalism~\cite{mori1965transport,Zwa73,ChHaKu00,ChHaKu02,hijon2009mori},
conditional expectations~\cite{Legoll:2010aa,Zhang2017,Legoll:2017aa},
averaging and homogenization~\cite{Pavliotis:2008aa}, Markov state models and
related techniques~\cite{Prinz2011c,Noe:2017aa}, and diffusion
maps~\cite{Rohrdanz:2011aa}. 

The starting point in the design of a coarse molecular model is the definition
of the variables. The choice of the coarse coordinates is usually made by
replacing a group of atoms by one effective particle, and is usually based on
physical and chemical intuition.  Because of the modularity of a protein
backbone or a DNA molecule, popular models coarse-grain a macromolecule to a
few interaction sites per residue or nucleotide, e.g., the $C_{\alpha}$ and
$C_{\beta}$ atoms for a protein \cite{ClementiJMB2000,MatysiakClementi_JMB04_Perturbation,Davtyan2012}. 
Alternative methods have been proposed to design coarse variables more
systematically \cite{Ponzoni20151516,sinitskiy2012optimal,BoninsegnaBanish2018}. 

In this study, we are concerned with the inference of governing equations on
the reduced set of variables, given that these descriptors have already been
selected, and that simulation data of the full system is available. 
Several methods to learn the parameters of an effective dynamics from data of the full
system have been proposed, most notably the force matching
scheme~\cite{izvekov2005multiscale,noid2008multiscale} and the relative
entropy method \cite{Shell2008} (the two
approaches are connected \cite{Rudzinski2011}). Sparse learning of dynamical equations has been studied in Refs. \cite{brunton2016discovering,boninsegna2018sparse}.
Most of the previous work was aimed at recovering correct
thermodynamics of the reduced system, that is, the distribution sampled by the
effective dynamics should equal the distribution of the projected original
process.
However, these methods do not determine the equations for a system's dynamical
evolution. That means that these methods may not be able to design
coarse-grained models that can reproduce molecular dynamical mechanisms (e.g.
large conformational changes or assembly mechanisms in protein systems).
Here, we shift the focus to designing coarse-grained models that reproduce slow
dynamical mechanisms of a fine-grained system,
that is, timescales, metastable states and transitions in between them.

In principle, if the dynamical equations of the fine-grained model are known,
the dynamics of the corresponding coarse-grained variables is given by the Mori-Zwanzig
projection formalism
\cite{mori1965transport,Zwa73,ChHaKu00,ChHaKu02}, which introduces a memory
term.  Even if the memory term can be simplified with the
assumption of a separation of timescales, the estimation of the quantities
involved in the Mori-Zwanzig approach is non trivial \cite{hijon2009mori}. However, here
we are not interested in reproducing all timescales of the system, but
just the slowest processes (e.g., conformational changes or assembly processes
in protein systems). This fact allows us to bypass the Mori-Zwanzig approach to directly
define an effective coarse-grained potential to satisfy this requirement.
We build on a framework for parameter estimation of stochastic dynamics
introduced in Refs. \cite{Crommelin:2006aa,Crommelin:2011aa}. It exploits that
slow dynamical processes are directly related to low-lying eigenvalues and
associated eigenfunctions of the generator of the
dynamics~\cite{Davies1982a,Davies:1982aa}. A
broad range of methods is available to approximate these spectral components
from simulation data of the full system
\cite{Deuflhard2000f,Prinz2011c,Noe2013c,Schutte:2013aa,Nuske2014a,mardt2018vampnets}. 
Hence, we argue that the same loss function can be used to learn coarse-grained
dynamics if the focus is on reproducing the slow kinetics, and call the
resulting framework the \textit{spectral matching} estimator. We derive the
optimization problems for two specific use cases of spectral matching: the
first is to recover an effective potential within an overdamped dynamics, the
second is to correct dynamics obtained from force matching by learning a
position-dependent diffusion. Applications to several toy systems and molecular
dynamics simulations of alanine dipeptide illustrate the capabilities of the
method and highlight practical details, especially the importance of
regularization.

\section{Theory}

\subsection{Full Space Dynamics}
We consider a stochastic process $X_t$ attaining values in $d$-dimensional
space $\mathbb{R}^d$, where $d$ is typically large. In the case of a molecular
system, $X_t$ represents the coordinates of the molecule at time $t$. We assume the process
solves a reversible stochastic differential equation (SDE):
\begin{eqnarray}
\label{eq:general_rev_process}
dX_t &=& \left[-A(X_t) \nabla_x F(X_t) + \nabla_x \cdot A(X_t)\right] dt + \sqrt{2} \sigma(X_t) dW_t.
\end{eqnarray}
In this formulation, $F$ is a scalar potential, while $A = \sigma \sigma^T$ is
the diffusion, which is a field of symmetric positive definite matrices. The
divergence $\nabla_x \cdot A$ is understood as applying the divergence operator
to each row of $A$. In the following, we will focus on $A$ instead of
$\sigma$ as all statistical properties of the process depend only on $A$.
Equation (\ref{eq:general_rev_process}) is the general form of a reversible
SDE, where reversibility holds with respect to the unique invariant density of
$X_t$, given by $\mu(x) \propto \exp(-F(x))$. A widely used example of Eq.
(\ref{eq:general_rev_process}) is the overdamped Langevin dynamics defined by a
potential energy function $V$ and a constant diffusion $a > 0$, which is
related to temperature $T$ and friction $\gamma$ by $a = \frac{k_B T}{\gamma}$:
\begin{eqnarray}
\label{eq:overdamped_langevin}
dX_t &=& -a \nabla V(X_t) dt + \sqrt{2 a}\, dW_t.
\end{eqnarray}
The generator associated to Eq. (\ref{eq:general_rev_process}) is the second order
differential operator:
\begin{eqnarray}
\label{eq:operator}
\mathcal{L}f &=& \left[-A \nabla_x F + \nabla_x \cdot A\right]\cdot \nabla_x f + A : \nabla^2_x f,
\end{eqnarray}
where the colon indicates the Frobenius inner product between matrices. The
non-negative operator $-\mathcal{L}$ is self-adjoint on the space $L^2_\mu$ of
functions square-integrable with respect to the weight function $\mu$. We
assume $-\mathcal{L}$ possesses a discrete set of increasing eigenvalues $0 =
\kappa_0 < \kappa_1 < \ldots$, with associated eigenfunctions $\psi_i,\,i=0, 1,
\ldots$. Each eigenvalue $\kappa_i$ corresponds to the
relaxation rate of a dynamical process in the system \cite{Pazy:1983aa}, with
\textit{characteristic (implied) timescale}:
\begin{eqnarray}
\label{eq:definition_timescale}
t_i &=& \frac{1}{\kappa_i}.
\end{eqnarray}
We are particularly interested in metastable processes; in terms of the
eigenspectrum of (\ref{eq:operator}) that means that there is a
cluster of $M$ eigenvalues $\kappa_1, \ldots, \kappa_M$ close to $\kappa_0 =
0$, separated from all higher eigenvalues by a spectral gap
\cite{Davies:1982aa,Davies1982a,Dellnitz:1999aa,Deuflhard2000f}. We can see
from Eq. (\ref{eq:definition_timescale}) that these low-lying eigenvalues
correspond to slow relaxation processes. We thus assume that there is a
separation of timescales between fast and slow processes in the molecular
system of interest.

\subsection{Reduced Dynamics}
In this study, we consider coarse graining of the system
(\ref{eq:general_rev_process}) by projecting the state space $\mathbb{R}^d$ into a
lower-dimensional space $\mathbb{R}^m, \, m\leq d$. Following  Refs
\cite{Legoll:2010aa,Zhang:2016aa}, this projection is realized by a
\textit{coarse-graining map} $\xi:\,\mathbb{R}^d \mapsto \mathbb{R}^m,\,
x\mapsto z=\xi(x)$. Our objective is to replace (\ref{eq:general_rev_process})
by another reversible SDE defined only on the lower-dimensional space
$\mathbb{R}^m$:
\begin{eqnarray}
\label{eq:reduced_rev_process}
dZ_t  &=& \left[-A^\xi(Z_t) \nabla_z F^\xi(Z_t) + \nabla_z \cdot A^\xi(Z_t)\right] dt + \sqrt{2} \sigma^\xi(Z_t) dW_t.
\end{eqnarray}
The effective potential $F^\xi$ and effective diffusion $A^\xi$ should meet the following two requirements: 
\begin{enumerate}
\item \textbf{Thermodynamic Consistency}: The invariant density $\nu =
\exp(-F^\xi)$ of (\ref{eq:reduced_rev_process}) should equal the integral of
the full state invariant measure $\mu$ over pre-images of $\xi$:
\begin{eqnarray}
\label{eq:thermodyn_consistency}
\nu(z) &=& \int_{\Sigma_z}\mu(x) \det \left[(\nabla_x \xi)^T \nabla_x \xi\right]^{-1/2}  \,\mathrm{d}x \\
\label{eq:definition_mu_z}
&=:& \int_{\Sigma_z}\,\mathrm{d}\mu_z(x).
\end{eqnarray}
Here $\Sigma_z = \{ x\in \mathbb{R}^d:\, \xi(x)=z \}$ is the pre-image of $z$
under the coarse-graining map, and $\nabla_x \xi \in \mathbb{R}^{d\times m}$ is
the Jacobian of $\xi$. Thus, the measure $\mathrm{d}\mu_z(x)$ defined in
(\ref{eq:definition_mu_z}) is the restriction of the full invariant measure to
$\Sigma_z$, multiplied by a correction factor accounting for the non-linear
change of variables. If $\nu$ satisfies Eq. (\ref{eq:thermodyn_consistency}),
the corresponding scalar potential is called the \textit{potential of mean
force (pmf)}:
\begin{eqnarray}
F^\xi_{mf}(z) = -\log \left[\int_{\Sigma_z}\,\mathrm{d}\mu_z(x)\right].
\end{eqnarray}
\item \textbf{Kinetic Consistency}: The coarse-grained dynamics should retain
the metastable part of the original dynamics (that is, the slow processes).
This requirement can be expressed using the generator $\mathcal{L}^\xi$ of
(\ref{eq:reduced_rev_process}):
\begin{eqnarray}
\mathcal{L}^\xi f &=& \left[-A^\xi \nabla_z F^\xi + \nabla_z \cdot A^\xi\right]\cdot \nabla_z f + A^\xi : \nabla^2_z f,
\end{eqnarray}
which is self-adjoint on the space $L^2_\nu$ of functions of $z$
square-integrable with respect to $\nu$. The leading (non-zero) eigenvalues
$\kappa^\xi_1, \ldots, \kappa^\xi_M$ and eigenfunctions $\psi^\xi_1,\ldots,\psi^\xi_M$ of $-\mathcal{L}^\xi$ should match the
corresponding eigenvalues and eigenfunctions of the original system, that is, they should satisfy 
$\kappa^\xi_i \approx \kappa_i$ and $\psi^\xi_i \approx \psi_i$, for $i = 1,\ldots, M$ (note that we always have
$\kappa^\xi_0 = 0 =\kappa_0$).
\end{enumerate}

\subsection{Force Matching}
The thermodynamic consistency can be enforced by \textit{force Matching}
\cite{izvekov2005multiscale,noid2008multiscale}, a
powerful technique to extract the potential of mean force from simulation data
of the full system. It is based on the fact that the gradient of the pmf solves
the following minimization problem
\cite{noid2008multiscale,ciccotti2008projection}:
\begin{eqnarray}
\label{eq:force_matching_problem}
\nabla_z F^\xi_{mf} &=& \mathrm{argmin}_{g\in L^2_\nu} \int_{\mathbb{R}^d} \|
g(\xi(x)) - F^\xi_{lmf}(x)\|^2\,\mathrm{d}\mu(x), \\
\label{eq:definition_lmf}
F^\xi_{lmf} &=& -\nabla_x F \cdot  G^\xi + \nabla_x \cdot G^\xi, \\
G^\xi &=& \nabla_x \xi \left[(\nabla_x \xi)^T \nabla_x \xi\right]^{-1},
\end{eqnarray}
where the minimization is over all square-integrable vector fields of the
reduced variables $z$. In (\ref{eq:definition_lmf}), $F$ is the scalar
potential from Eq. (\ref{eq:general_rev_process}), while $F^\xi_{lmf}$ is called
\textit{local mean force}. If sufficient simulation data $X_{t_1}, \ldots,
X_{t_K}$ of the full system is available, Eq. (\ref{eq:force_matching_problem}) can
be replaced by a data-based regression:
\begin{eqnarray}
\nabla_z F^\xi_{mf} &=& \mathrm{argmin}_{g\in L^2_\nu}  \frac{1}{K}\sum_{k=1}^K \| g(\xi(X_{t_k})) - F^\xi_{lmf}(X_{t_k})\|^2.
\end{eqnarray}
We note that force matching can also be applied if the original process is full
Langevin dynamics in position and momentum space, and $\xi$ is a transformation
of the position space only.
One of the difficulties in the practical application of this method has been
that, in general, a coarse-grained potential satisfying the thermodynamic consistency
includes many-body terms that are not easily modeled in the energy functions. 
Recently, machine learning methods have been used to alleviate this problem
\cite{Deshmukh2018,ZhangHan2018_CG,wang2018machine,Chan2019}.

\subsection{Spectral Matching}
In order to achieve kinetic consistency, we build on an idea presented in Refs.
\cite{Crommelin:2006aa,Crommelin:2011aa} for parameter estimation in stochastic
dynamics like Eq. (\ref{eq:general_rev_process}). Given the leading spectral
components $\kappa_i,\, \psi_i,\, i=1,\ldots, M$ of the full system, a
parametric model $\mathcal{L}_\theta$ for the generator, and a set of
\textit{test functions} $f_j,\, j=1,\ldots,P$, in \cite{Crommelin:2006aa,Crommelin:2011aa} it was 
suggested to minimize the discrepancy in the leading eigenvalue equations in a
weak sense:
\begin{eqnarray}
\label{eq:spectral_matching_idea}
\theta^* &=& \mathrm{argmin}_\theta\, \frac{1}{2}\sum_{i=1}^M\sum_{j=1}^P 
\langle \mathcal{L}_\theta \psi_i + \kappa_i \psi_i, f_j \rangle_\mu^2.
\end{eqnarray}
Application of this idea to the coarse-grained case is based on two insights.
First, if thermodynamic consistency holds, inner products in coarse-grained
space can be estimated using data of the full system, as for any $f,g \in
L^2_\nu$, we have \cite{Zhang:2016aa}:
\begin{eqnarray}
\label{eq:expectations_coarea}
\langle f, g \rangle_\nu &=& \langle f\circ \xi, g\circ \xi \rangle_\mu.
\end{eqnarray}
Second, we have recently shown \cite{Nueske2019} that it is indeed possible to
find a kinetically consistent model if the full eigenfunctions can be well
approximated by functions of the reduced variables, i.e. if
\begin{eqnarray}
\label{eq:approximate_eigenfunctions}
\psi_i(x) &\approx & \tilde{\psi}^\xi_i(\xi(x)) = \tilde{\psi}^\xi_i(z)
\end{eqnarray}
holds for $i = 1,\ldots, M$. Spectral approximation and validation techniques
like TICA \cite{Perez-Hernandez:2013aa}, Markov state models
\cite{Deuflhard2000f,Prinz2011c,Schutte:2013aa,Bowman2014b}, the Variational
Approach \cite{Noe2013c,Nuske2014a}, or VAMPnets \cite{mardt2018vampnets} (see
below) can be used to obtain the approximations $\tilde{\psi}^\xi_i$, and to
verify that (\ref{eq:approximate_eigenfunctions}) holds.

Thus, given approximate spectral components $\tilde{\kappa}^\xi_i,\,
\tilde{\psi}^\xi_i(z)$, a set of test functions $f_j(z)$ on the coarse-grained
space, a parametric model $\mathcal{L}^\xi_\theta$ for the coarse-grained
generator, and data sampling the full space distribution $\mu$, we can combine
Eqs. (\ref{eq:spectral_matching_idea} - \ref{eq:approximate_eigenfunctions}) to
define the \textit{spectral matching} estimator:
\begin{eqnarray}
\theta^* &=& \mathrm{argmin}_\theta \, E_1(\theta) \\
E_1(\theta) &=& \frac{1}{2}\sum_{i=1}^M\sum_{j=1}^P \langle
\mathcal{L}^\xi_\theta \tilde{\psi}^\xi_i + \tilde{\kappa}^\xi_i
\tilde{\psi}^\xi_i, f_j \rangle_\nu^2 \\
&=& \frac{1}{2}\sum_{i=1}^M\sum_{j=1}^P \langle \mathcal{L}^\xi_\theta
\tilde{\psi}^\xi_i \circ \xi + \tilde{\kappa}^\xi_i \tilde{\psi}^\xi_i \circ
\xi, f_j \circ \xi \rangle_\mu^2.
\end{eqnarray}
Alternatively, we can also have the model generator act on the test functions,
which yields:
\begin{eqnarray}
\label{eq:spectralmatching}
\theta^* &=& \mathrm{argmin}_\theta \, E_2(\theta) \\
E_2(\theta) &=&  \frac{1}{2}\sum_{i=1}^M\sum_{j=1}^P \left [ \langle
\tilde{\psi}^\xi_i, \mathcal{L}^\xi_\theta f_j \rangle_\nu +
\tilde{\kappa}^\xi_i \langle \tilde{\psi}^\xi_i, f_j \rangle_\nu \right]^2.
\end{eqnarray}
Whether or not $E_1$ and $E_2$ are identical depends on whether the model
generator $\mathcal{L}^\xi_\theta$ is symmetric w.r.t. the inner product
$\langle \cdot\,  , \, \cdot \rangle_\nu$. In other words, the two are identical if the model dynamics are reversible with respect to the measure $\nu$, or yet in other words, if the scalar potential $F^\xi$ of the model is induced by $\nu$ as $F^\xi = \exp(-\nu)$.

\section{Methods}

In this section, we discuss two specific use-cases for the spectral matching
estimator, as well as details of the practical implementation of the method.

\subsection{Estimation of a Scalar Potential}

The first use-case arises from the assumption that the effective dynamics can be
modeled using overdamped Langevin dynamics in a scalar potential
$F^\xi_\theta$, using a fixed constant diffusion $a > 0$. The resulting model
generator is
\begin{eqnarray}
\label{eq:Langevin}
\mathcal{L}^\xi_\theta f &=& -a\nabla_z F^\xi_\theta \cdot \nabla_z f + a \Delta_z f.
\end{eqnarray}
In particular, if the model is a linear expansion into given basis functions,
i.e. $F^\xi_\theta = \sum_{n=1}^N w_n g_n$, $\theta = (w_1, \ldots, w_N)\in
\mathbb{R}^N$, spectral matching becomes a linear regression. In order to avoid
the computation of second order derivatives for the eigenfunctions, we apply
the model generator to the test functions (Eq. \ref{eq:spectralmatching}). We also include the zeroth spectral
pair, as the corresponding matrix entries are non-zero, to obtain a regression
matrix $X \in \mathbb{R}^{(M+1)P\times N}$, and a data vector $y\in
\mathbb{R}^{(M+1)P}$:
\begin{eqnarray}
\label{eq:spectral_matching_F_linear}
E_2(w) &=& \frac{1}{2}\| Xw - y \|^2, \\
X_{i,j;n} &=& -\langle \tilde{\psi}^\xi_i, \nabla_z g_n \cdot \nabla_z f_j \rangle_\nu, \\
y_{i,j} &=& - \langle \tilde{\psi}^\xi_i, \Delta_z f_j + \tilde{\kappa}^\xi_i f_j \rangle_\nu.
\end{eqnarray}
We have found  that regression (\ref{eq:spectral_matching_F_linear}) is often
ill-conditioned and requires regularization. Below, we will use elastic net
regularization \cite{zou2005regularization} with parameters $\alpha,\,\rho$,
that is:
\begin{eqnarray}
\label{eq:ridge_regression}
E_2^\alpha(w) &=& \frac{1}{2}\| Xw - y \|^2 + \alpha\rho \|w\|_1 + \frac{1}{2}\alpha(1-\rho) \|w\|_2^2.
\end{eqnarray}

\subsection{Estimation of a Diffusion}

Next, we consider the situation where an estimate of a thermodynamically
consistent scalar potential $F^\xi_{mf}$ is already available, for example by
application of the force matching scheme. As a result, any parametric model
$A^\xi_\theta$ for the diffusion results in a symmetric model generator, and
the parameter dependent term in the loss function can be evaluated as:
\begin{eqnarray}
\langle \mathcal{L}^\xi_\theta \tilde{\psi}^\xi_i, f_j\rangle_\nu &=& - \int
A^\xi_\theta \cdot \nabla_z \tilde{\psi}^\xi_i \cdot \nabla_z f_j \,\mathrm{d}\nu \\
&=& \langle \tilde{\psi}^\xi_i, \mathcal{L}^\xi_\theta f_j\rangle_\nu,
\end{eqnarray}
provided that $A^\xi_\theta$ is always symmetric. This formulation only
requires first order derivatives, but we need to estimate those derivatives for
the eigenfunctions. As a special case, we focus on a linear expansion into
scalar multiples of the identity matrix, i.e. $A^\xi_\theta =
\left[\sum_{n=1}^N w_n g_n \right] \mathrm{Id}, \, \theta = (w_1, \ldots,
w_N)\in \mathbb{R}^N$. Again, spectral matching leads to a linear regression:
\begin{eqnarray}
\label{eq:spectral_matching_A_linear}
w^* &=& \mathrm{argmin}_w \, \frac{1}{2}\| Xw - y \|^2, \\
X_{i,j;n} &=& -\langle g_n \nabla_z \tilde{\psi}^\xi_i \cdot \nabla_z f_j \rangle_\nu, \\
y_{i,j} &=& - \tilde{\kappa}^\xi_i \langle \tilde{\psi}^\xi_i,  f_j \rangle_\nu.
\end{eqnarray}
Positive definiteness of the diffusion is easily enforced in this setting, by
requiring positivity of the scalar pre-factor at all data points.  In fact, we
will  restrict the diffusion to satisfy pre-selected upper and lower bounds $0
\leq a_{min}(z) \leq a_{max}(z)$, which can in principle be position dependent.
If $K$ data points $Z_{t_1} = \xi(X_{t_1}), \ldots, Z_{t_K} = \xi(X_{t_K})$ are
given, we add linear inequality constraints:
\begin{eqnarray}
\label{eq:constraints_A}
a_{max}(Z_{t_k}) &\geq \sum_{n=1}^N w_n g_n(Z_{t_k}) &\geq a_{min}(Z_{t_k}) \geq 0,
\end{eqnarray}
for all $k= 1,\ldots, K$. The full optimization problem
(\ref{eq:spectral_matching_A_linear}) then turns into a quadratic programming
problem.

\subsection{Variational Approach}
\label{subsec:vac}
The spectral matching approach discussed above requires an estimate of the
eigenfunctions and eigenvalues of the original system.
In order to compute approximations of a system's dominant eigenfunctions
$\psi_i$ and corresponding eigenvalues $\kappa_i$, we make use of the
\textit{Variational Approach to Conformational Dynamics} (VAC)
\cite{Noe2013c,Nuske2014a}, which is a data-driven method to represent the
dominant eigenfunctions from a given library of basis functions. Below, the
library will either consist of Gaussian functions or of piecewise constant
functions, in which case the method is called a \textit{Markov state model}
\cite{Deuflhard2000f,Prinz2011c,Schutte:2013aa,Bowman2014b}. Both techniques
are employed to obtain the approximations $\tilde{\psi}^\xi_i, \,
\tilde{\kappa}^\xi_i$ required as inputs to the spectral matching, and also to
analyze simulation data of the effective dynamics obtained from spectral
matching. In order to extract a decomposition of state space into metastable
sets, based on the dominant eigenfunctions, we use the PCCA algorithm
\cite{Deuflhard:2005aa,Roblitz:2013aa}. The software implementation we use is
the pyEmma package \cite{scherer2015pyemma}.

\section{Results}

\subsection{Three Well Potential}
\label{subsec:three_well}
\begin{figure}
\begin{centering}
\includegraphics{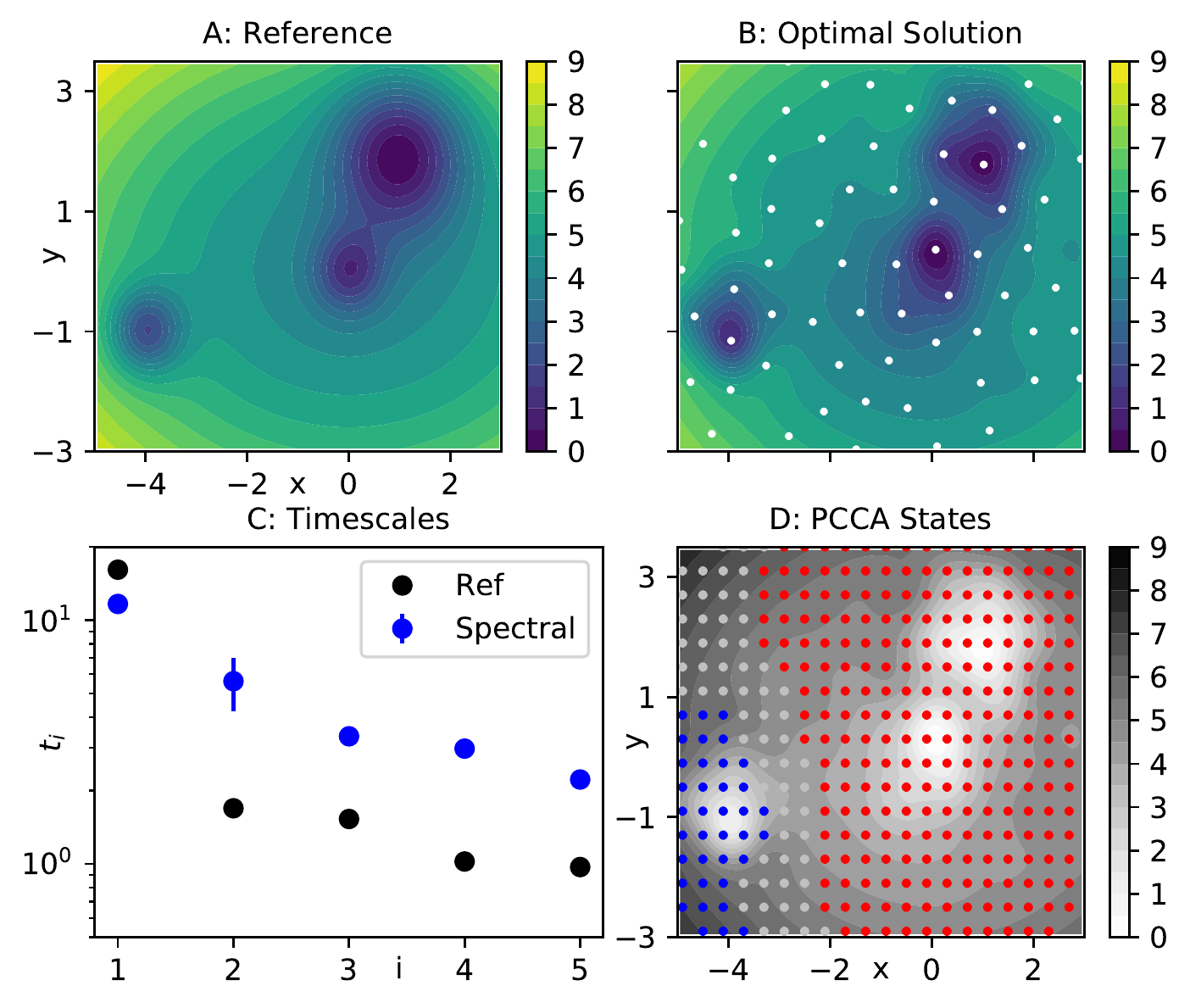}
\par\end{centering}
\caption{Application of spectral matching with a linear model, cf Eq.
(\ref{eq:spectral_matching_F_linear}), to a two-dimensional test system. A) The
reference potential Eq. (\ref{eq:2d_three_well_energy}). B) Optimized potential
found by spectral matching. White dots indicate the locations of 82
centers used to define the Gaussian basis and test functions. C) Comparison of
the first five implied timescales of the reference dynamics (black) and the
corresponding timescales of the learned potential (blue). The spectral matching
was designed to reproduce the slowest timescale, as indeed appears in the
figure. D) Metastable
decomposition obtained by applying 2-state PCCA to an MSM of the learned
dynamics. States are assigned to a macrostate if their membership is larger
than 0.6. Silver dots represent transition states that cannot be assigned in
this way. \label{fig:three_well}}
\end{figure}

We first illustrate the idea of spectral matching to model the overdamped
Langevin dynamics (Eq. \ref{eq:Langevin}) in a two-dimensional toy potential. The energy function is
given as a sum of three Gaussians and a
harmonic confinement, that is:
\begin{eqnarray}
\label{eq:2d_three_well_energy}
V(x, y) &=& \sum_{q=1}^3 a_q \exp(-\frac{1}{2s_q^2}((x-m_q^x)^2 + (y-m_q^y)^2) + a_4 x^2 + a_5 y^2.
\end{eqnarray}
The actual values of the parameters are $(a_1, a_2, a_3, a_4, a_5) = (-3, -5,
-4, 0.1, 0.2)$, $(m_1^x, m_1^y) = (0, 0)$, $(m_2^x, m_2^y) = (1, 2)$, $(m_3^x,
m_3^y) = (-4, -1)$, and $(s_1, s_2, s_3) = (0.5, \sqrt{5/6}, 0.5)$; a contour of
the energy is shown in Fig. \ref{fig:three_well} A. The diffusion constant is
$a = 1$. We generate a long equilibrium simulation of these dynamics, at
integration time step $\Delta_t = 10^{-4}$, spanning $K = 10^7$ time steps.

We use a Markov state model based on a grid discretization of the
two-dimensional state space to compute the slow eigenfunctions and eigenvalues
of the system. The slowest process corresponds to the transition out of the
shallow minimum on the left into the center, it occurs at a timescale $t_1
\approx 16$. The rest of the spectrum is separated from this process and there
is no other distinct slow motion to be identified, see the black dots in Fig.
\ref{fig:three_well} D. We thus set $M = 2$ (that is, we use only the first
non-trivial eigenfunction) and extract the approximate eigenfunction
$\tilde{\psi}^\xi_1$ from the MSM.

We attempt to reproduce the energy $V$ in two-dimensional space from data, while not
applying any model reduction. We employ spectral matching and a linear model in
combination with the elastic net as in Eq. (\ref{eq:ridge_regression}). A
regular space clustering of the data at minimum distance 0.8 helps us define a
set of 82 spherical Gaussian functions centered at the cluster centers, as shown
in Fig. \ref{fig:three_well} B. These centers serve to define the test set,
while the basis set consists of the same Gaussians plus the quadratic functions
$x^2,\, y^2$ ($P = 82,\, N = 84$). We use uniform widths $\sigma_{base}$ and
$\sigma_{test}$ for all Gaussians in the basis set and the test set,
respectively. We fix $\sigma_{test} = 0.8$, while $\sigma_{base}$ and both
elastic net parameters $\alpha,\, \rho$ are treated as hyper-parameters. They
are determined by 3-fold cross-validation (CV) for all triples of the form
$(\sigma_{base}, \alpha, \rho) \in \{0.2, 0.4, 0.6, 0.8, 1.0\} \times
\{10^{-6}, 10^{-4}, 10^{-2}, 10^0 \} \times \{0.5, 0.8, 0.9, 0.95, 1.0 \}$.
The optimal parameter set is $(\sigma_{base}, \alpha, \rho) = (0.4, 10^{-4}, 0.5)$, 
the corresponding solution is displayed in Fig. \ref{fig:three_well} B and
agrees well with the original energy landscape in panel A.

Using the same simulation parameters as for the original system, we generate a
realization of the dynamics defined by the optimized potential. An MSM analysis
confirms that the slowest timescale is indeed well reproduced by those dynamics, see
Fig. \ref{fig:three_well} C. As only the first non-trivial eigenfunction was
used in the spectral matching, we do not expect to reproduce additional
timescales besides the slowest one.
Moreover, application of 2-state PCCA to this
Markov model also shows that the decomposition of state space into metastable
sets is the same as for the original potential, as displayed in panel D of Fig
\ref{fig:three_well}.

\subsection{Three Well Potential with Roughness}
\begin{figure}
\begin{centering}
\includegraphics[scale=0.45]{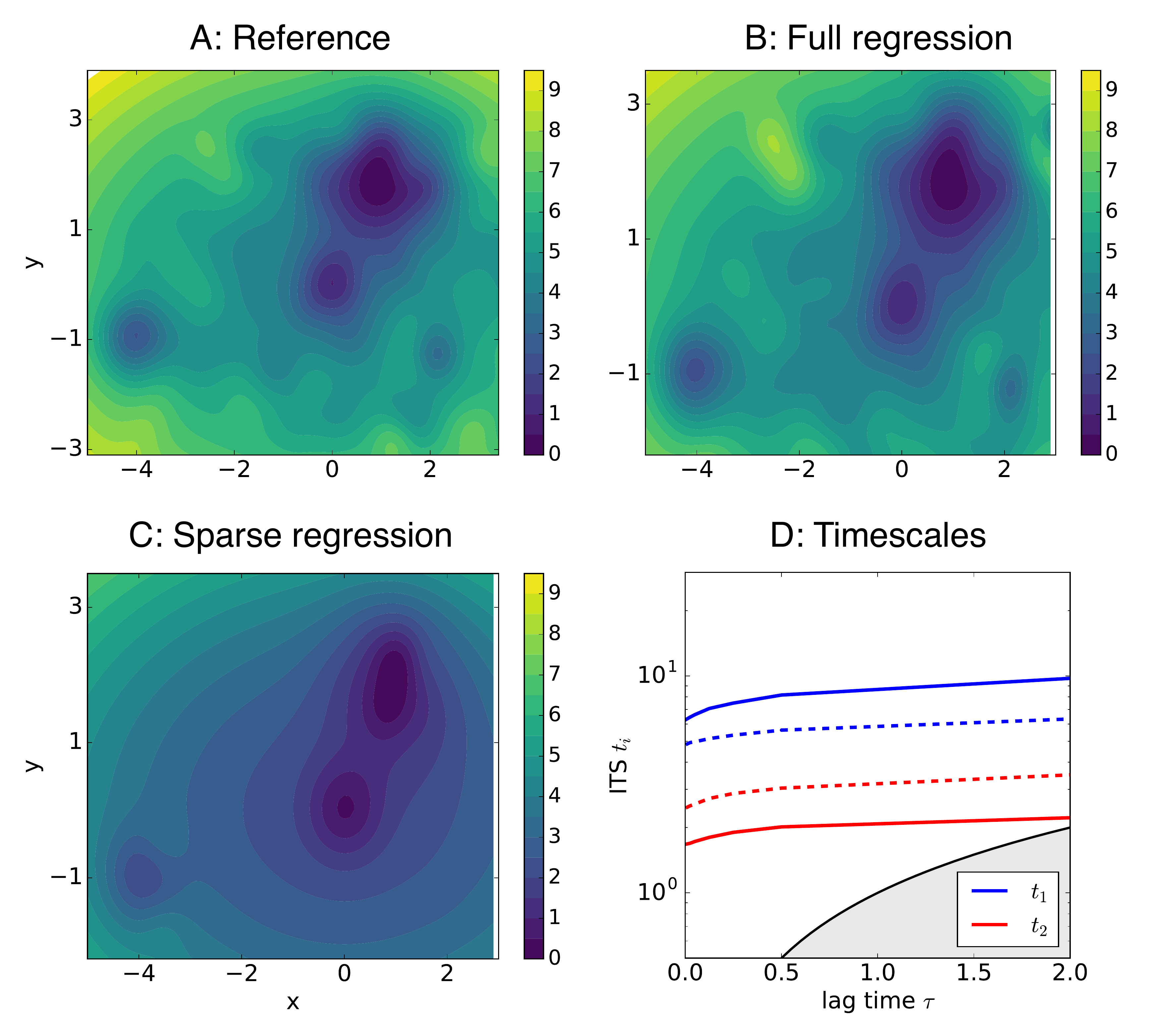}
\par\end{centering}

\caption{Application of spectral matching with a linear model, cf Eq.
(\ref{eq:spectral_matching_F_linear}), to a rugged potential. A) The reference
potential is shown and was built by adding $100$ small amplitude Gaussians to
Eq. (\ref{eq:2d_three_well_energy}), which are responsible for the ripples,
compare with Fig. \ref{fig:three_well} A. B) Resulting
potential obtained from running standard linear regression. C) Optimal solution
obtained after optimizing regurarizing  hyper-parameters $\rho, \,\alpha$ by
cross-validation. D) First two implied timescales $t_1$ and $t_2$ as a function of lag time
$\tau$ estimated from simulations of the reference (blue) and the learned
potential (red).\label{fig:three_well_perturbed}}
\end{figure}

Here, the performance of the spectral matching is tested on a perturbed version
$V_{per}$ of the three well potential investigated in section
\ref{subsec:three_well}, to which $100$ small amplitude Gaussians were added;
i.e., 
\begin{equation}
\label{eq:2d_three_well_energy_perturbed}
V_{per}(x, y) = V(x, y) + \sum_{q=1}^{100} w_q \exp \lbrace{-\frac{(x - x^0_q)^2 + (y - y^0_q)^2}{2\sigma^2} \rbrace},
\end{equation}
with $\sigma = 0.2$, and $(x^0_q, y^0_q)$ indicating the center of the $q$-th
perturbing Gaussian;  a contour plot of such a potential is shown in Fig.
\ref{fig:three_well_perturbed} A. The diffusion constant is $a = 1$. We
generate a long equilibrium simulation of these dynamics, at integration time
step $\Delta_t = 10^{-4}$, spanning $K = 6 \cdot 10^7$ time steps.

The dominant timescales $t_i$ and eigenfunctions $\tilde{\psi}_i$ were
numerically approximated using the VAC and a basis of 196 Gaussian features.

Spectral matching was applied to the data set, as exactly the same  functions
entering the linear combination Eq. (\ref{eq:2d_three_well_energy_perturbed})
were employed both as basis and test functions in the procedure. Eventually,
the full set of $105$ coefficients $\{a_i\}_1^5 \bigcup \{w_q\}_{1}^{100}$ is
approximated and compared with the exact result. Regression was solved using
standard implementations, with and without elastic net regularization.
The first two non-trivial eigenpairs were used in the spectral matching (that
is, $M=3$ in Eq. \ref{eq:spectralmatching}).

The non-regularized solution is shown in Fig. \ref{fig:three_well_perturbed} B
and almost perfectly reproduces the exact potential, panel A. The regularized
solution is shown in panel C, and the optimal regularization hyper-parameters
$\rho, \alpha = (0.5, 2 \cdot 10^{-4})$ were identified by running 3-fold CV
on all tuples of the form $(\rho, \alpha) \in \{0.4, 0.5, 0.6\} \times
\{10^{-4}, 2\cdot10^{-4}, 3\cdot10^{-4} \}$. The solution is very sparse, only $17$
coefficients (out of the full set of $105$) have non-zero values. The position
of the three main energy wells is correctly identified, which correlate with
the system's slowest dynamics, while most of roughness of the potential
is washed out. Regularization appears to be a physically meaningful procedure, since it
systematically filters perturbations, i.e., rugged local minima, out.

However, the wells are shallower than expected, compare with Fig.
\ref{fig:three_well_perturbed} A. In order to investigate if this observation
has any effect, a long equilibrium trajectory was generated by integrating
overdamped Langevin dynamics, using the optimal solution as potential energy
($\Delta_t = 10^{-4}, \, a = 1$), and the associated timescales were estimated
using VAC in the same setup as before. Results are shown in Fig.
\ref{fig:three_well_perturbed} D, where the resulting two slowest timescales (red)
are compared to the reference ones (blue): the processes occur on very similar timescales.

\subsection{Recovery of Slow Kinetics by Non-Constant Diffusion}
\begin{figure}
\begin{centering}
\includegraphics{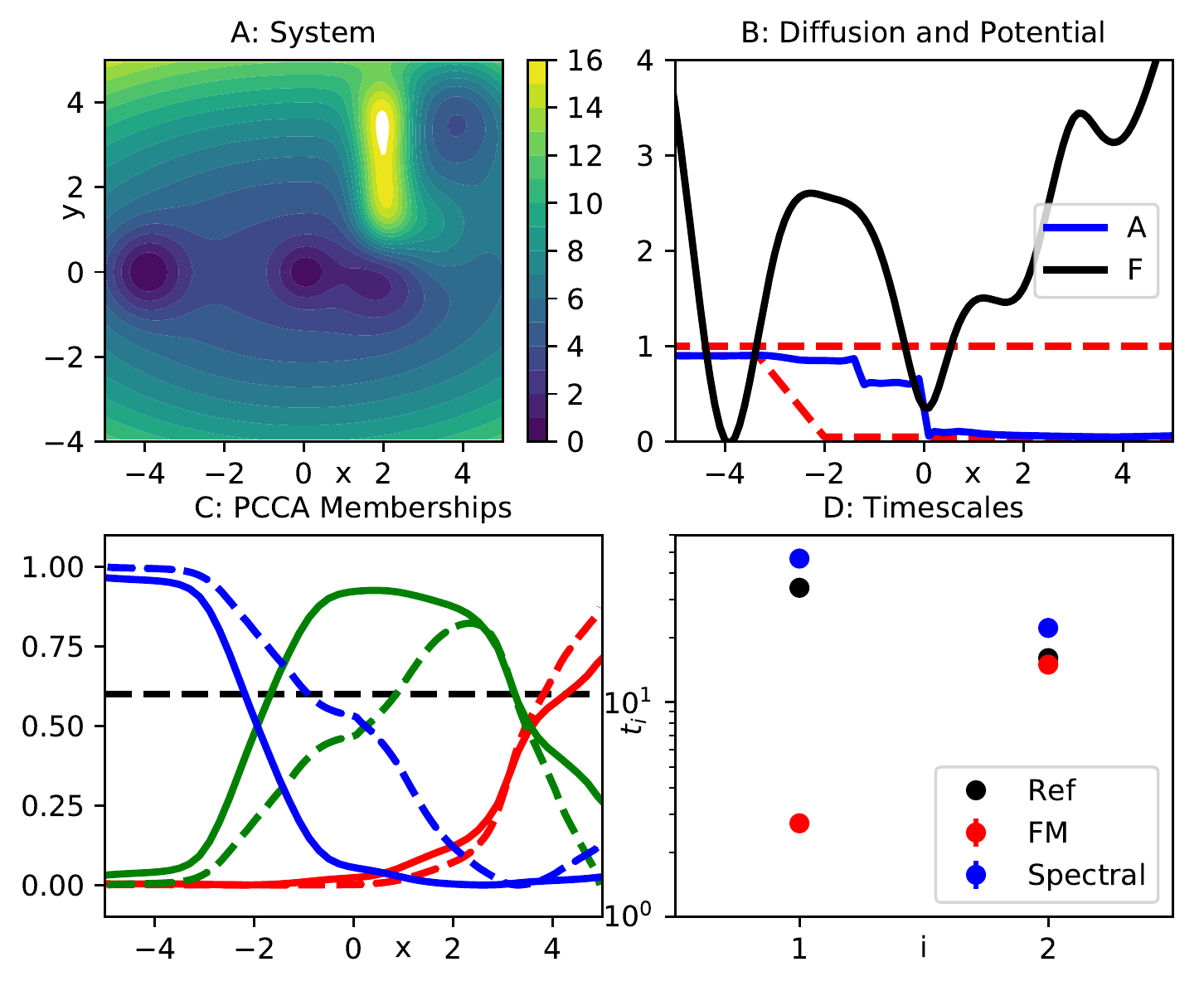}
\par\end{centering}

\caption{A) Results for suboptimal projection of a two-dimensional toy
potential. A) Full potential.  B) Estimates of the potential of mean force
(black) and position-dependent diffusion (blue). Red dashed lines correspond to
the lower and upper bounds $a_{min},\,a_{max}$ for the diffusion. C) Metastable
memberships extracted by running three-state PCCA on Markov state models of the
full dynamics after projection onto $x$ (solid lines) and of the effective
simulations (dashed lines). The black line at membership equal to 0.6 is given
as a visual aid to determine metastable sets. D) Comparison of slowest two
timescales obtained from the full simulation (black), simulations of the force
matching potential (red), and simulations of the combined dynamics
(\ref{eq:reduced_rev_process}) using the force matching potential and the
optimized diffusion (blue). Note that we have exchanged the order of the force
matching timescales to ensure all timescales for index $i$ correspond to the
same slow process. Errorbars were calculated using a Bayesian MSM.
\label{fig:1d_diffusion}}
\end{figure}

Next, we demonstrate that spectral matching can be used to recover the slow
kinetics of a projected system with the aid of a non-constant diffusion.
We consider a modified three-well system (Fig. \ref{fig:1d_diffusion} A), where
the locations of the minima have been changed, and additional peaks have been
added to the landscape. The slowest process is now the transition out of the
shallow minimum on the right. An equilibrium
simulation of 10 million frames at integration time step $\Delta_t = 10^{-3}$
serves as the reference data set.

We estimate the reduced dynamics along the first coordinate $\xi(x, y) = x$. We
apply force matching using a basis of 71 Gaussian functions centered at grid
spacing 0.2 between $x = -7.0$ and $x = 7.0$. A uniform width is selected by
means of CV from the set $\sigma \in \{0.1, 0.2, 0.4, 0.6, 0.8 \}$. The
estimated potential of mean force for the optimal value $\sigma = 0.2$ is
represented by the black line in Fig \ref{fig:1d_diffusion} B.

We generate simulation data of overdamped Langevin dynamics in the
potential of mean force, at diffusion $a = 1$ (using the same integration time
step and number of frames as for the full system). An MSM analysis of these
data shows that the slowest timescale is decreased by a factor ten, while the
next timescale is almost unaffected by the projection, see Fig.
\ref{fig:1d_diffusion} D. Thus, the order of the two timescales is reversed,
and the slow kinetics cannot be recovered by simply re-scaling the effective
diffusion constant $a$.

Spectral matching in the form (\ref{eq:spectral_matching_A_linear}) is applied
to estimate a position-dependent diffusion. Approximate rates
$\tilde{\kappa}_i^\xi,\, i=1, 2$ are extracted from a Markov state model of the
full process. As the approximate eigenfunctions $\tilde{\psi}^\xi_i$ need to be
functions of $z$ in Eq. (\ref{eq:spectral_matching_A_linear}), we build another
MSM along $x$ alone and extract those approximate eigenfunctions. Even though
this MSM does not yield accurate estimates of both timescales, the
corresponding eigenfunctions still capture the structure of both slow
transitions correctly, as we can see by looking at the corresponding PCCA
memberships (solid lines in Fig. \ref{fig:1d_diffusion} C). The test set is
chosen as the same basis set used for force matching. As a basis set, we choose
a set of $N$ piecewise constant functions, where the corresponding discrete
sets are obtained by partitioning the interval $[-6, 6]$ into $N$ equal-sized
parts. The optimal choice for $N$ is again obtained by 3-fold cross validation
on the hyper-parameter set $N \in \{5, 10, 20, 30, 40 \}$.

It is important to solve Eq. (\ref{eq:spectral_matching_A_linear})
subject to position dependent lower bounds $a_{min}$. Based on our analysis of
the full data and the force matching simulation, we require $a_{min} = 0.9$ ,
that is, close to one, in the left metastable set, while we set $a_{min} = 0.1$
everywhere else. We also use $a_{max} = 1.0$ as a global upper bound. The
resulting optimal diffusion is indicated by the blue line in Fig.
\ref{fig:1d_diffusion} B. This solution tends to slow down the transition out
of the rightmost metastable state, while leaving the timescale of transition
out of the center state unchanged. After running the effective dynamics
(\ref{eq:reduced_rev_process}) for 10 million steps at $\Delta_t = 10^{-3}$,
and analyzing these data by an MSM, we find that both timescales are indeed
approximately restored, see Fig. \ref{fig:1d_diffusion} D. These effective
dynamics are more diffusive than the original system, as is reflected in the
corresponding PCCA memberships being less crisp (dashed lines in Fig.
\ref{fig:1d_diffusion} C). The change in the dynamics to restore the correct
timescales perturbs the position of the metastable sets. However, the three
metastable sets of the original dynamics are approximately retained.

\subsection{Alanine Dipeptide}
Finally, we study model reduction of a small molecular system, alanine
dipeptide, which has served as a test case for numerous studies in recent
years. The data set at our disposal is the same that was used in Ref.
\cite{wang2018machine}, please see Ref. \cite{Nuske2017} for the detailed
simulation setup. It comprises one million frames of Langevin dynamics in
explicit water saved every one picosecond.

\begin{figure}
\begin{centering}
\includegraphics{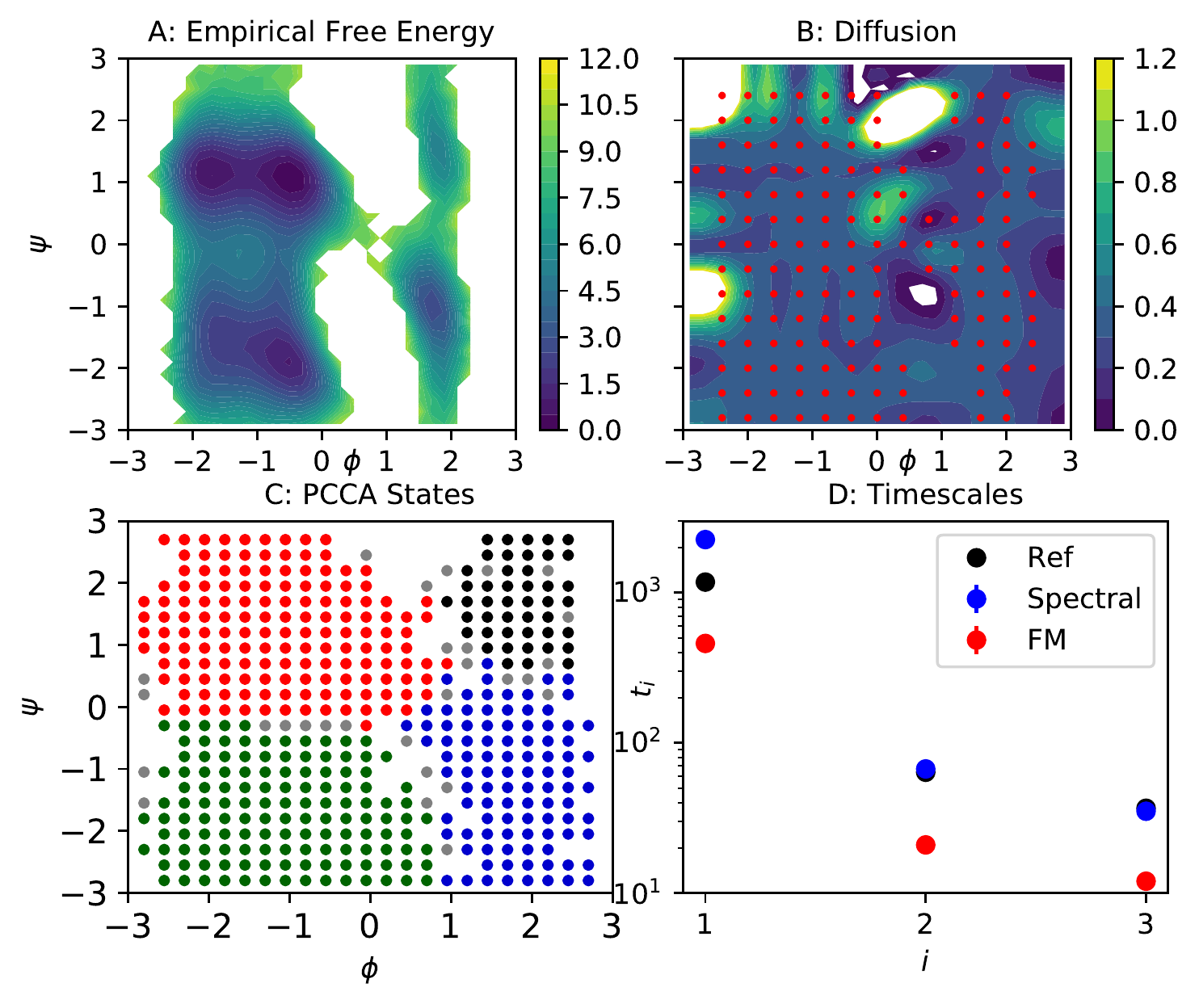}
\par\end{centering}

\caption{Results for coarse-graining of alanine dipeptide in the space of its
backbone dihedrals $\phi,\, \psi$. A) Empirical free energy in dihedral space
from original MD simulation. B) Optimized diffusion. Red markers indicate
centers of the Gaussian functions defining the basis and test set. C)
Metastable states as identified by four-state PCCA on an MSM of the effective
dynamics. A state is assigned to a PCCA state if its degree of membership is
larger than 0.6. Gray dots represent transition states that cannot be assigned
to a macrostate in this way.  D) Comparison of three slowest implied timescales
obtained from the reference
dynamics (black), simulations of the force matching potential (red), and
simulations combining the optimal diffusion from C) and the force matching
potential (blue). Errorbars were generated using a Bayesian Markov state model.
\label{fig:ala2_results}}
\end{figure}

It is well known that the system's metastable processes are effectively
functions of its two backbone dihedral angles $\phi, \, \psi$. Therefore, we
study reduction of this system into the two-dimensional space of those dihedral
angles. Figure \ref{fig:ala2_results} A shows the empirical free energy in this
space. It presents a more challenging example than the toy systems, as the
coarse graining map is non-linear, and the metastability is much more
pronounced in the sense that there are large unsampled areas and we only have
very few samples in the transition regions. To simplify matters, we shift both
dihedrals in order to eliminate almost all periodicity from this
representation, and use non-periodic basis functions in what follows.

To obtain the approximate eigenfunctions $\tilde{\psi}^\xi_i$ and eigenvalues
$\tilde{\kappa}^\xi_i$, we apply the VAC with 225 spherical Gaussian functions
centered on a regular grid between -2.8 and +2.8 at a distance of 0.4 in each
direction. Their widths are uniformly set to 0.3. We retain the first three
non-trivial slow eigenfunctions, and the corresponding implied timescales are
$t_1 \approx 1.2\, \mathrm{ns}$, $t_2 \approx 63\, \mathrm{ps}$, and $t_3 \approx
36\,\mathrm{ps}$. The corresponding transitions in dihedral space occur,
respectively, between the left and right half of the plane, between the two
minima on the left, and between the two shallow minima on the right.

We use spherical Gaussian functions to represent the force matching potential
and to serve as basis and test functions in the spectral matching. The centers
are placed on a regular grid with grid spacing 0.4 in all directions, but we
remove centers in unpopulated areas of the state space. As a result, a set of
152 centers for Gaussian basis functions is obtained. For force matching, we
employ 3-fold cross validation to determine a uniform width out of the
parameter set $\sigma \in \{0.2, 0.3, 0.4, 0.5, 0.6, 0.7, 0.8\}$, to yield an
optimal value $\sigma = 0.6$. To ensure that the resulting potential is
confining, we evaluate the solution on a fine grid, and replace its values at
unsampled grid points by a function that grows quadratically with the distance
to the sampled area. A two-dimensional spline is fitted to this data to yield
the final approximation.

To evaluate the kinetic properties of the force matching model, we generate a
set of 100 overdamped Langevin simulations of $15\,\mathrm{ns}$ length at
constant diffusion $a = 1$. By comparing the three slowest timescales of these
dynamics to the reference values (red and black dots in Fig.
\ref{fig:ala2_results} D), we find that the force matching dynamics are
uniformly accelerating the kinetics by about a factor three.

For spectral matching, we use the optimal Gaussian functions from force
matching to serve as test functions, and also use the same centers to define
the basis set. These centers are indicated by red markers in Fig.
\ref{fig:ala2_results} B. We globally set $a_{min} = 0.3,\, a_{max} = 1.0$, and
apply 3-fold CV to determine the widths of the basis set, where $\sigma_{base}
\in \{0.4, 0.45, 0.5, 0.55, 0.6\}$. The solution of Eq.
(\ref{eq:spectral_matching_A_linear}) for the optimal parameter value
$\sigma_{base} = 0.55$ is shown in Fig. \ref{fig:ala2_results} B. The resulting
diffusion is mostly constant at a value close to 0.3 throughout most of the
domain, except for some peaks in boundary or unsampled regions. By generating
another set of 100 simulations of $15\,\mathrm{ns}$ simulation time each, we
confirm that all three timescales are corrected by applying the optimized
diffusion on top of the scalar potential, see Fig. \ref{fig:ala2_results} D,
blue dots. By applying four-state PCCA to the effective simulations, we also
verify the  corresponding metastable sets to coincide with those of the
original dynamics, see Fig. \ref{fig:ala2_results} C. Thus, the spectral
matching enables us to find a (mostly constant) diffusion term in order to
correct the coarse-grained system's kinetic properties. 

As a further consistency check, we also apply spectral matching with just a
single basis function, which is the constant, while leaving all other settings
unchanged. The resulting expansion coefficient, which is nothing more than an
effective diffusion constant, also equals 0.3.

\section{Conclusions}

We have introduced spectral matching as a method to define effective
coarse-grained models reproducing the dynamics of a
fine-grained system.  Spectral matching can be used as stand alone or in
combination with the existing force-matching method. While force matching
enforces thermodynamic consistency, spectral matching enforces kinetic consistency.
The goal of spectral matching is to retain slow processes of the original dynamics, while
faster motions are considered less relevant and may be lost in the process. For
two specific settings, we have presented the resulting data-based regression
problems that follow from spectral matching. The first setting is the
estimation of an effective potential for an overdamped dynamics, the second is
concerned with learning an effective diffusion to correct the kinetics induced
by force matching. We have demonstrated by several examples that spectral
matching can be used to learn governing equations which retain the slow part of
the dynamics. We found that suitable regularization is vital to the success of
the method.

Several questions are left open for future work. While we were able to use relatively simple basis sets here, the application of the method to large molecular systems will require the use of more powerful model classes. Using deep learning, or other state of the art machine learning techniques,  in conjunction with spectral matching is the topic of ongoing research. A better theoretical understanding of the method is also required. For instance, the limiting behavior of spectral matching if the test and basis sets become exhaustive needs to be addressed, especially in the case where a potential different from the pmf is estimated by spectral matching. Moreover, we found in many cases that multiple different solutions with almost equivalent kinetic properties can be found by spectral matching. A systematic treatment of this phenomenon will also follow in future work.

\bibliography{Library}
\end{document}